\documentclass[aps,prl,twocolumn,showpacs,superscriptaddress,groupedaddress,footinbib]{revtex4-1}  % for review and submission
\usepackage{graphicx}  % needed for figures
\usepackage{dcolumn}   % needed for some tables
\usepackage{bm}        % for math
\usepackage{amssymb}   % for math

\begin{document}

% The following information is for internal review, please remove them for submission
%\widetext
%\leftline{Version xx as of \today}
%\leftline{Primary authors: Joe E. Physics}
%\leftline{To be submitted to (PRL, PRD-RC, PRD, PLB; choose one.)}
%\leftline{Comment to {\tt d0-run2eb-nnn@fnal.gov} by xxx, yyy}
%\centerline{\em D\O\ INTERNAL DOCUMENT -- NOT FOR PUBLIC DISTRIBUTION}

% the following line is for submission, including submission to the arXiv!!
%\hspace{5.2in} \mbox{Fermilab-Pub-04/xxx-E}

\title{Creep events and creep noise in gravitational-wave interferometers: basic formalism and stationary limit}
\author{Yuri Levin$^1,^2$}
\email{yuri.levin@monash.edu.au}
\affiliation{$^1$ Monash Center for Astrophysics, Monash University, Clayton, VIC 3800, Australia}
\affiliation{$^2$ Leiden Observatory, Leiden University, Niels Bohrweg 2, Leiden, the Netherlands}

\begin{abstract} \noindent
In gravitational-wave interferometers, test masses are suspended on thin fibers which experience considerable tension stress. Sudden microscopic stress release in a suspension fiber, 
which I call a 'creep event',  would  excite motion of the test mass that would be
 coupled to the interferometer's readout. The random test-mass motion
 due to a time-sequence of creep events is referred to as 'creep noise'. In this paper I present an
  elasto-dynamic calculation
 for the test-mass motion due to a creep event. 
 I show that within a simple suspension model, the main coupling to the optical readout  occurs via a combination of a ``dc" horizontal displacement of the test mass, and excitation of the violin and pendulum modes, and not, as was thought previously, via lengthening of the fiber.
When the creep events occur sufficiently frequently and their statistics is time-independent, the creep
noise can be well-approximated by a stationary Gaussian random process. I derive the functional form
of the creep noise spectral density in this limit, with the restrictive assumption that the creep events are statistically independent from each other.

%On a technical side, I prove a reciprocity theorem in linear elastodynamics that was instrumental 
%for my calculations. The theorem may useful for other
%applications. 
\end{abstract}
%\pacs{interferometers, gravitational waves, elasticity theory}
%\clearpage
%\begin{multicols}{2}
%
\maketitle
\section{I. Introduction}

Gravitational-wave interferometers like Laser Interferometric Gravitational-wave Observatory (LIGO) 
in the United States of America \cite{abbott2009},  VIRGO \cite{acernese} in Europe, and their smaller counterparts GEO600 in Germany \cite{luck} and TAMA in Japan \cite{arai}, are using
super-precise opto-mechanical measurements to search for astrophysical gravitational waves. 
After several years of taking scientific data, LIGO and VIRGO are
currently being upgraded with improved  instrumentation and should again become operational in
2015 \cite{harry2010}, \cite{accadia2011}. 
LIGO Science Collaboration (LSC) and the VIRGO community are projecting  \cite{abadie} that 
with the upgraded technology, both interferometers will soon be measuring multiple coalescences of 
relativistic compact objects (neutron stars and black holes). These projections are based in part on the theoretical predictions for spectral density of the interferometers' noise. It is thought that the random 
processes that contribute most of the noise, i.e. the seismic shaking of the suspensions \cite{coughlin}, the thermo-mechanical and thermo-refractive
fluctuations of the mirror surface \cite{harrybook}, 
and the quantum-mechanical fluctuations of the light-field coupled
to the test-mass motion \cite{KLMTV}, \cite{buonannochen} are well understood \cite{footnote1}.

One of the dangerous unknowns for the advanced gravitational-wave interferometers is a non-Gaussian noise from a superposition of transient events in the instrument. In this paper I concentrate 
on the creep noise, which is caused by a superposition of the sudden localized 
tension stress releases (creep
events) in suspension fibers and their end attachments. 
It has been thought that a creep event would couple to the interferometer's readout via lengthening of
the fiber \cite{cagnoli}. Specifically, it was argued that because of the Earth' curvature, the laser beam was not strictly perpendicular to the suspension fiber, and thus the fiber's lengthening would result in some test-mass displacement 
along the beam. In this paper I show that this coupling, while present, is not dominant, at least for a simple model where the fiber  is represented by a cylinder with constant radius. Instead, a creep
event couples to the interferometer's output predominantly through excitation of the pendulum and 
violin modes of the suspension; this coupling is explicitly calculated in this work. 

The fact that creep
events couple to the transverse vibrational modes of the system is in agreement with the experiment
of \cite{ageev} who find a substantial excess noise in the transverse motion of a tungsten wire stretched to 20\%
of the break-up stress.  Similar excess noise in steel wires was observed by \cite{bilenko1}. However, the results in \cite{ageev} were not  confirmed by
\cite{gretarsson} who did not observe any excess noise in the motion of the stressed tungsten wire.  Moreover, it is far from obvious that the processes responsible for the creep events in metallic fibers \cite{desalvo} will be operating in the fused silica suspension fibers such as the ones that are currently used in GEO600 and that will be used in the advanced LIGO, VIRGO, and KAGRA suspensions \cite{aston}. Two experiments with fused silica fibers have been performed by \cite{bilenko2} and \cite{gretarsson};
in both experiments no excess noise was discovered near violin resonant frequency of the fiber. In a more recent work \cite{sorazu}, the motion of a test mass was monitored in GEO600, where the fused silica
suspension fibers were used. The motion near the violin-mode frequency was entirely consistent with
that of the thermally excited violin mode. Therefore, currently there is no experimental evidence that the creep excess noise in the future advanced gravitational-wave interferometers will pose a serious problem. However,   there are at least two reasons to keep investigating the creep noise: (1) the measurements 
in \cite{bilenko2}, \cite{gretarsson}, and \cite{sorazu} have all been performed at  frequencies from
several hundreds to thousands of Hz, where the noise of ground-based interferometers is strongly dominated by the quantum shot noise, while the danger from creep noise exists at much lower frequencies, in the same region of tens of Hz where the shot noise is unimportant, and (2) the main 
source of creep noise may well be not inside the fused silica suspension  fibers, but inside other 
carrying parts of the system like the bond between the test masses and the ``ears" that are supporting
them (Riccardo DeSalvo and Norna Robertson, private communications). It is thus important to understand how a creep
event inside the suspension couples to the horizontal motion of the test-mass, as well as the frequency
dependence of the noise generated by a multitude of the creep events. This paper lays a theoretical
foundation for addressing these issues.

The plan of the paper is as follows. In section 2, I present a convenient reciprocity relationship for 
linear elasto-dynamic systems. In section 3, I use this relationship to derive the interferometer's response
to a creep event, as a function of location of the stress release in the fiber. In section 4, I derive the
functional form of the creep noise spectral density, in the limit where the creep noise can be treated
as a stationary Gaussian random process. A brief discussion of the future work is presented in 
section 5. 

\section{II. Elastodynamics and reciprocity theorem}
During the creep event, the stress changes suddenly in some small volume of the fiber. But how does this
affect the motion of the test mass? At first glance, this seems like a formidable problem in 
elastodynamics. However, it turns out that solving the reciprocal problem is sufficient. Namely, 
one should in a mental experiment apply a sudden force to the test mass and compute the motion of the
fiber at the location where the creep event originated. The solution of the reciprocal
problem leads directly to the solution of the original problem. 
Reciprocity relations have been thoroughly studied in elastodynamics; see e.g.~\cite{achenbach}
for a comprehensive review. Here I will use the following formulation of
 the reciprocity theorem:

Consider an elastodynamic system initially at rest, that is being driven by
a distributed force, with the force-per-volume given by
\begin{equation}
\vec{F}(\vec{r},t)=\vec{f}(\vec{r}) \chi(t),
\label{force1}
\end{equation}
where $\chi(t)$ is some function that is non-zero only \newline for $t>0$. Consider also a readout variable
\begin{equation}
{X}(t)=\int d^3r \vec{g}(\vec{r})\cdot \vec{\xi}(\vec{r},t),
\label{displacement}
\end{equation}
where $\vec{\xi}(\vec{r},t)$ is the displacement from rest at location $\vec{r}$ and time $t$. Both the
applied forces and displacement are assumed to be small, so that a linear approximation of elastodynamics holds. The reciprocity theorem states that if in the pair of  
Equations~(\ref{force1}) and (\ref{displacement}) 
the form-factors $\vec{f}(\vec{r})$ and $\vec{g}(\vec{r})$ are interchanged then the
readout variable ${X}(t)$ remains the same. In other words, the input-output dynamical system is
invariant with respect to $\vec{f}$ and $\vec{g}$ interchange, with $\chi(t)$ being the input and $X(t)$ being the output.

The proof of the theorem is  as follows. Let $\vec{\xi}_n(\vec{r})$ be the normal modes of the system, with
proper angular frequencies $\omega_n$. A displacement field $\vec{\xi}$ can then be decomposed 
into a series
\begin{equation}
\vec{\xi}(\vec{r},t)=\Sigma_n a_n(t) \vec{\xi}_n(\vec{r}).
\label{decomposition}
\end{equation}
The mode amplitudes $a_n(t)$ are the new dynamical coordinates. The Lagrangian  of the unforced
system is given by
\begin{equation}
L_0={1\over 2}\Sigma_n m_n \left[\dot{a}_n^2-\omega_n^2 a_n^2\right],
\label{Lfree}
\end{equation}
where $m_n$ is the effective mass of the $n$'th mode.
External forcing of Eq.~(\ref{force1}) is introduced via an additional interaction Langangian term
\begin{equation}
L_ {\rm int}=\int d^3r \vec{F}(\vec{r},t)\cdot\vec{\xi}(\vec{r},t),
\end{equation}
which, in terms of the coordinates $a_n$ can be rewritten as
\begin{equation}   
L_{\rm int}=\Sigma_n a_n f_n \chi(t).
\end{equation}
Here $f_n$ are constants given by
\begin{equation}
f_n=\int d^3r \vec{\xi}_n(\vec{r})\cdot \vec{f}(\vec{r}).
\end{equation}
The full Langrangian allows us to immediately obtain the equations of motion:
\begin{equation}
{d^2a_n\over dt^2}+\omega_n^2 a_n=\chi(t){f_n\over m_n}.
\label{eqmotion}
\end{equation}
Therefore, 
\begin{equation}
a_n(t)={f_n\over m_n}\chi_n(t),
\label{solution1}
\end{equation}
where $\chi_n(t)$ is the solution to the forced harmonic oscillator problem
\begin{equation}
{d^2\chi_n\over dt^2}+\omega_n^2\chi_n=\chi(t)
\end{equation}
with the initial condition $\chi_n(0)=\dot{\chi}_n(0)=0$. The readout variable in Eq.~(\ref{displacement})
can then be written as 
\begin{equation}
X(t)=\Sigma_n{g_nf_n\over m_n}\chi_n(t),
\label{displacement1}
\end{equation}
where $g_n$ is defined similarly to $f_n$: 
\begin{equation}
g_n=\int d^3r\vec{\xi}_n(\vec{r})\cdot \vec{g}(\vec{r}).
\end{equation}
The readout variable $X(t)$ is invariant with respect to the interchange of $\vec{f}$ and $\vec{g}$. Q.~E.~D.

\section{III. Test-mass response to a single creep event}
\subsection{General considerations}
A creep event happens when a minute section of the suspension fiber refuses to support its full share
of the tension stress. What exactly happens microscopically is poorly known, but a simple model
will suffice for modelling of the elastodynamical behavior. Let us assume that a small fiber 
element of volume $V$ suddenly does not support any elastic stress $T_{ij}$. 
%which is the dominant component
%of the stress tensor in the fiber (it is assumed that the vertical axis $z$ is aligned with the fiber at rest).
%One way to visualize this is to recall that $T_{zz}$ represents the $z$-directed flux of the $z$-component
%of the momentum density. Thus our little volume $V$ of the fiber is no longer stransporting this 
%momentum-density flux, it it has to flow around $V$ thus increasing the stress elsewhere. The cancellation of the $T_{zz}$ inside $V$ can be achieved by adding the force at the boundary,
I now consider a  slightly-reduced elastic system,  namely the original one with the small volume
element $V$ 
taken out. This slightly-reduced system experiences a sudden force applied to the boundary of
the volume element $V$, 
so that the boundary surface element $\vec{dS}$, assumed to be directed outside of the volume, 
experiences the force 
\begin{equation}
\vec{dF}=-T_{ij} (\vec{dS}\cdot \vec{e}_j) \vec{e}_i,
\label{compensator}
\end{equation}
where $\vec{e}_i$ are the unit vectors along the coordinate axes, and the summation over the dummy 
indices is assumed.

I would like to evaluate the test-mass displacement $X(t)$ under the action of the
force in Eq.~(\ref{compensator}) that is switched on at $t=0$ (this situations is somewhat similar 
physically to the excitation of magnetar motion as a result of sudden reconfiguration
of the magnetosphere during a giant magnetar flare; see \cite{levin1}). 
By the reciprocity theorem from
the previous section, this is equivalent to acting with the suddenly switched on  force on the test mass, directed along the laser beam:
\begin{equation}
{F}_{\rm test mass}(t)=\Theta(t)
\label{Ftheta}
\end{equation}
where $\Theta(t)$ is the Heavyside function \cite{footnote2}. One then has to find the response of the slightly-reduced elastic subsystem to this force, and in particular that of the reciprocal readout
variable ${X}_{\rm readout}(t)$ that is dictated by the functional form of the force in Eq.~(\ref{compensator})
\begin{equation}
{X}_{\rm readout}(t)=-\int T_{ij} \xi_i \vec{dS}\cdot \vec{e}_j,
\label{Xreadout0}
\end{equation}
where the integration domain is the boundary of the volume $V$. It is  obvious that for sufficiently
small \cite{footnote3} volume $V$ the response of the slightly-reduced system is the same as that of the
full system, and from hereon I shall make no distinction between the two.

By Gauss' theorem, for small $V$ the above equation can be written as 
\begin{equation}
X_{\rm readout}(t)=-VT_{ij}{\partial\xi_i\over \partial x_j}
\label{Xreadout}
\end{equation}
where the strain $\partial\xi_i/\partial x_j$ is evaluated at the location of the creep event.
To sum up: by finding the response of $X_{\rm readout}$ from Eq.~(\ref{Xreadout}) to the
force $F_{\rm test mass}=\Theta(t)$ applied at the test mass along the direction of the laser beam,
one finds the test-mass displacement in response to the creep event.

It is convenient and instructive to work in the Fourier domain
\begin{equation}
F_{\rm test mass}(t)=\int _{-\infty}^{\infty} d\omega F_{\rm test mass}(\omega) e^{i\omega t}.
\label{Fourier1}
\end{equation}
For the force given by Eq.~(\ref{Ftheta}), the force Fourier component is given by
\begin{equation}
F_{\rm test mass}(\omega) = \lim_{\epsilon\rightarrow 0+}\left\{{1\over 2i\pi(\omega-
i\epsilon)}\right\},
\label{fourierforce}
\end{equation}
where positive $\epsilon$ serves to avoid the singularity  in Eq.~(\ref{Fourier1}); the limit 
$\epsilon\rightarrow 0-$ would give ${F}_{\rm test mass}(t)=\Theta(-t)$.

The test mass horizontal displacement $X_{\rm testmass}$ under the action of the applied force is given by
\begin{equation}
X_{\rm testmass}(\omega) ={F_{\rm testmass}(\omega)\over M}Z[(\omega-i\gamma(\omega)],
\label{testmassresponse}  
\end{equation}
where the mechanical impedance $Z(\omega)$, as derived in the Appendix,
 is given by \cite{footnote4} 
\begin{equation}
Z[\omega]={1\over
\pi{\omega\over \omega_s}\cot\left[\pi{\omega\over \omega_s}\right]\omega_p^2-\omega^2}.
\label{impedance}
\end{equation}
Here, $M$ is the mass of the test mass, $\omega_p=\sqrt{g/l}$ is the pendulum angular frequency of the
test mass,  $\omega_s=\pi\sqrt{M/(Nm)}\omega_p$ is the fundamental violin mode angular
frequency 
measured when the test-mass is fixed in space, and $l$, $m$, and $N$ are the length, the mass, and 
the number of the strings on which the test mass is suspended.  The small positive $\gamma(\omega)\ll\omega$ inserted
into Eq.~(\ref{testmassresponse}) represent damping. Mathematically, this displaces the poles of 
$Z(\omega)$ into the upper half of the complex $\omega$-plane \cite{footnote5}. These poles represent the frequencies
of normal modes of the suspension; their imaginary parts equal the rate of exponential decay of their
amplitudes. The actual values of $\gamma(\omega)$ are only important near the normal-mode
frequencies and can be measured experimentally.

If no damping is present, the poles of $Z[\omega]$, i.e.~the normal-mode frequencies, are given approximately by
\begin{equation}
\omega_0\simeq\pm\omega_p=\pm\sqrt{g/l}
\label{pendulum}
\end{equation}
for the pendulum mode, and
\begin{equation}
\omega_{vj}\simeq\pm\left[j\omega_s+(-1)^j{\omega_p^2\over j\omega_s}\right]
\label{violin}
\end{equation}
for the $j=1, 2, ...$ violin modes. Let us introduce non-zero $\gamma_p$ and $\gamma_{vj}$ which are the damping rates of
the pendulum and violin modes, respectively. 
The impedance can be expanded as follows:
\begin{eqnarray}
Z(\omega-i\gamma)&\simeq&{1\over \omega_p^2-(\omega-i\gamma_p)^2}+\label{impedance1}\\
                &   &\Sigma_{j=1}^{\infty}{2\omega_p^2\over \omega_{vj}^2}
                       {1\over \omega_{vj}^2-(\omega-i\gamma_{vj})^2}.\nonumber
\end{eqnarray}
Substituting Eqs~(\ref{impedance}) and (\ref{fourierforce}) into Eq.~(\ref{testmassresponse}),
and evaluating the inverse Fourier transform, I get
\begin{eqnarray}
X_{\rm testmass}(t)&=&{1\over M\omega_p^2}\left[1-\cos(\omega_pt)e^{-\gamma_pt}\right]+\label{xtestmass1}\\
& &\Sigma_{j=1}^{\infty}{2\omega_p^2\over M\omega_{vj}^4}\left[1-\cos(\omega_{vj}t)e^{-\gamma_{vj}
t}\right].\nonumber
\end{eqnarray}
As expected, the system's response to a sudden force is a constant displacement added to damped
oscillations due to excited pendulum and violin modes.

Lets take stock, and sum up what has been done so far.
In a reciprocal problem, one has to find out the induced shear at the location of the creep-event source
when a sudden force $F(t)=\Theta(t)$ is applied to the test mass. In this subsection I have done 
part of the problem, i.e.~I found the test-mass displacement under the action of the said force.
To make further progress, I need to choose a particular model for the suspension fiber
itself. In the next subsection I  consider one of the particular cases that can be dealt with analytically.
The treatment of complicated geometries is left for future work.

\subsection{Example: creep event in a cylindrical fiber with constant cross-section}
Let us consider in detail the case where the creep events occur inside a cyllindrical
fiber of constant crossection that is rigidly attached at the top to a suspension isolation plate and  
at the bottom to the test mass. It is assumed here that the allowed test-mass motion is a parallel translation but not rotation, as is the case when four suspension fibers are used.
The dominant part of the stress in the fiber is 
\begin{equation}
T_{zz}=-{Mg\over N\pi r^2},
\label{dominantstress}
\end{equation}
where $N$ is the number of suspension fibers and $r$ is the radius of the fibers' horizontal crossection.
The readout variable in Eq.~(\ref{Xreadout}) is given by
\begin{equation}
X_{\rm readout}={MgV\over N\pi r^2}{\partial\xi_z\over \partial z},
\label{Xreadout2}
\end{equation}
where $z$ is the vertical coordinate along the fiber. I choose $z=0$ at the fiber's top and $z=l$ at the fiber's bottom.

Let $x_c, y_c, z$ be the spatial coordinates of the source of a creep event, where $x_c, y_c=0$ corresponds to the location of fiber's axis at $z$, and $x_c$ is measured along the laser beam direction.
The vertical strain induced by the fiber's motion is given by (see, e.g., Chapter 11 of \cite{blandford})
\begin{equation}
{\partial\xi_z\over\partial z}=-x_c{\partial^2 \eta\over\partial z^2},
\label{deformation}
\end{equation}
 where $\eta(z)$ be horizontal
displacement of the fiber. 
Therefore, the readout variable is
\begin{equation}
X_{\rm readout}=-{MgVx_c\over N\pi r^2}{\partial^2 \eta\over\partial z^2}.
\label{Xreadout3}
\end{equation}
Let us now find the fiber motion $\eta(z,t)$.
Its dynamical equation of motion is given by (see, e.g.,
chapter 12 of \cite{blandford} or \cite{landau})
\begin{equation}
{\partial^2\eta\over \partial t^2}=c_s^2\left[{\partial^2\eta\over \partial z^2}-\lambda^2
{\partial^4\eta\over \partial z^4}\right].
\label{fibermotion}
\end{equation}
Here $c_s=\sqrt{Mgl/(Nm)}$ is the velocity of the tension wave in a fiber, and $\lambda$ is the characteristic bending length
given by 
\begin{equation}
\lambda={1\over 2}\sqrt{\pi E N\over Mg}r^2={1\over 2}(\xi_{0z,z})^{-1/2}r.
\label{lambda}
\end{equation}
Here $r$ is the radius of the fiber, $E$ is the Young modulus, and $\xi_{0z,z}$ is the intitial stretch
factor of the fiber under the loading force $Mg/N$ of the test mass. Advanced LIGO will use the fused-silica fibers with the following parameters: $r=2\times 10^{-4}m$, $M=40$kg, $E=72$GPa, $l=0.6$m, and 
$N=4$ fibers per test mass. This parameters produce the bending length $\lambda\simeq 0.001\hbox{m}\ll l$.

The periodic solutions of the homogenious equation (\ref{fibermotion}) can be written as 
\begin{equation}
\eta(z,t)\propto e^{i\omega t}e^{pz},
\label{trialsolution}
\end{equation}
where
\begin{equation}
p^2={1\pm\sqrt{1+4\left({\omega\lambda\over c_s}\right)^2}\over 2\lambda^2}.
\label{pvalue1}
\end{equation}
For frequencies of interest, $\omega\lambda/c_s\ll 1$, and thus the solutions feature two physically distinct branches, $p=\pm 1/\lambda$ and $p=\pm i\omega/c_s$. The former branch represents evanescent quasi-static
bending perturbations that will be large near the fiber's attachment points, while the latter represent
tension waves in a fiber. The boundary conditions $\eta(0)=\eta^{\prime}(0)=\eta^{\prime}(l)$ and 
$\eta(l)=X_{\rm testmass}$, together with $\lambda\ll l$, determine the full solution for the fiber:

\begin{figure*}
  \includegraphics[scale=0.7]{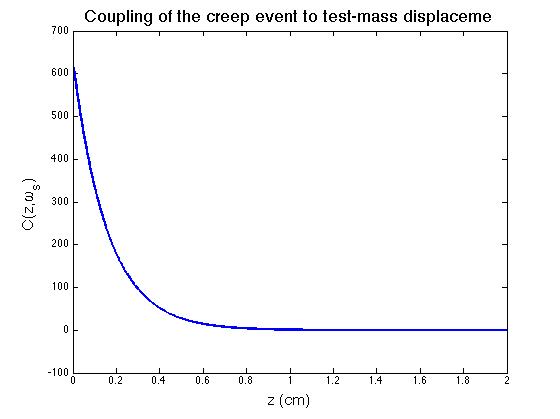}
\caption{\label{fig:epsart} The quantity $C(z,\omega)$ that characterizes the coupling strength of the creep event to the test-mass horizontal 
displacement, is plotted as a function of the creep-event's distance from the top of the fiber, at a frequency of the fundamental violin mode. The coupling is strongly peaked within the bending length $\lambda$ from the top. Similar ``bending" peak occurs at the bottom of the suspension fiber (not
shown). The parameters for the plot are those for advanced LIGO suspension fused silica fibers: $E=72$GPa, $r=2\times 10^{-4}$m, $l=0.6$m, $M=40$kg, $N=4$.}
\end{figure*}

\begin{equation}
\eta(z,t)=Be^{i\omega t}[\eta_{\rm bend}(z)+\eta_{\rm wave}(z)],
\label{yzt1}
\end{equation}
where 
\begin{equation}
\eta_{\rm bend}(z)=k\lambda \left\{e^{-{z\over\lambda}}-[\cos(kl)+k\lambda\sin(kl)]e^{z-l\over\lambda}\right\},
\label{ybend}
\end{equation}
and
\begin{equation}
\eta_{\rm wave}(z)=\sin(kz)-k\lambda\cos(kz).
\label{ywave}
\end{equation}
Here $k=\omega/c_s$, and the amplitude $B$  is given by
\begin{eqnarray}
B&\simeq & X_{\rm testmass}(\omega)/\sin(kl)\nonumber\\
&=&X_{\rm testmass}(\omega)/\sin(\pi\omega/\omega_s).
\label{B}
\end{eqnarray}
Therefore, in the Fourier domain the readout variable is given by

%Let $x_c, y_c, z$ be the spacial coordinates of the source of a creep event, where $x_c, y_c=0$ corresponds to the location of fiber's axis at $z$, and $x_c$ is measured along the laser beam direction.
%The deformation induced by the fiber's motion is given by
%\begin{equation}
%{\partial\xi_z\over\partial z}=-x_c{\partial^2 \eta\over\partial z^2}.
%\label{deformation}
%\end{equation}
%Collecting Eqs~(\ref{deformation}), (\ref{ybend}), (\ref{ywave}), (\ref{fourierforce}), and (\ref{impedance}),
%and noting $T_{zz}=Mg/(N\pi r^2)$, we now get an expression for the reciprocal readout variable in Eq.~(\ref{Xreadout}):
\begin{eqnarray}
X_{\rm readout}(\omega)&=&-\lim_{\epsilon\rightarrow 0+}{1\over 2\pi i (\omega-i\epsilon)}{gVx_c\over
N\pi r^2}\times\nonumber\\
&  &{1\over \pi{\omega\over \omega_s}\cos\left[\pi{\omega\over\omega_s}\right]\omega_p^2-\omega^2\sin\left[\pi{\omega\over
\omega_s}\right]}\times\nonumber\\
& &\left[\lambda^{-2} \eta_{\rm bend}(z,\omega)-k^2\eta_{\rm wave}(z,\omega)\right],\label{readout}
\end{eqnarray}
where $\eta_{\rm bend}(z,\omega)$ and $\eta_{\rm wave}(z,\omega)$ are given by Eqs.~(\ref{ybend}) and
(\ref{ywave}), respectively.
In the time-domain, putting in all the damping terms, I get
\begin{eqnarray}
X_{\rm readout}(t)&=&C_p(x_c, z)\left[1-\cos(\omega_p t)e^{-\gamma_pt}\right]+\label{xreadout}\\
                            & &\Sigma_{j=1}^{\infty}C_{vj}(x_c,z)\left[1-\cos(\omega_{vj} t)e^{-\gamma_{vj}t}\right]
                            \nonumber
\end{eqnarray}
where
\begin{equation}
C_p(x_c,z)=-{gVx_c\over N\pi r^2\omega_p^2}\left({e^{-{z\over \lambda}}-e^{z-l\over\lambda}\over\lambda l}-
{\omega_p^2\over \omega_s^2}{\pi^2\over l^2}{z\over l}\right),
\label{Cpz}
\end{equation}
and
\begin{eqnarray}
C_{vj}(x_c,z)&=&{2gVx_c(-1)^{j+1}\over N\pi r^2\omega_{vj}^2}\times\nonumber\\
& &\left\{{\pi j\over \lambda l}\left[e^{-{z\over l}}-(-1)^j e^{z-l\over \lambda}\right]-\right.\label{Cvjz}\\
  & &\left.\left({\pi j\over l}\right)^2
\sin\left(\pi j z/l\right)\right\}.\nonumber
\end{eqnarray}
In the expressions above I used  $\lambda\ll l$ and $\omega_p\ll\omega_s$. I remind the reader that
$x_c, z$ are the coordinates of the location of the creep event with the effective volume $V$, and that
Eq.~(\ref{xreadout}) gives the test-mass response to such an event.  The geometric nature of the 
prefactors in Eqs (\ref{Cpz}) and (\ref{Cvjz}) is apparent once one recalls $g/\omega_p^2\simeq l$, and
$g/\omega_{vj}^2\simeq [Nm/M](\pi j)^{-2} l$.

It is instructive to compute a numerical example. In the expression (\ref{xreadout}) above, consider
values $x_c=r$,
$V=\hbox{nm}^3$, and $z=0$ (i.e., a formation of a nanometer-size hole at the top edge of the fiber).
The displacement that one then gets at a pendulum frequency is or order $10^{-21}$m.

It is worthwhile to have another look at the right-hand side of Eq.~(\ref{readout}). The 
part of the equation in square brackets, 
\begin{equation}
C(z,\omega)=\lambda^{-2}\eta_{\rm bend}(z,\omega)-k^2\eta_{\rm wave}(z,\omega)
\label{C}
\end{equation}
determines the $z$-dependence  of the coupling of the creep event to the horizontal motion
of the test mass.  The function $C(z)$ is plotted in Fig.~1, for $\omega=\omega_s$ (i.e., the fundamental violin mode). Fiducial parameters that
were used in making the plot are
specified in the figure's header. The function peaks very strongly within $\lambda$
from the attachment ends of
the fiber; there $C(z)\sim k/\lambda$ is dominated by the $\eta_{\rm bend}$. Away from the 
attachment points, the coupling is dominated by the $\eta_{\rm wave}$ part of the solution
and $C(z)\sim k^2$. 
 
 In this subsection's model the creep events are assumed to be triggered  homogeneously in the suspension fibers \cite{footnote6}. Thus the creep events have only $\sim\lambda/l$ chance to be triggered
 within $\lambda$ from the attachment points. However, they have 
 individually much larger impact [by a factor of $1/(k\lambda)$] on the test-mass motion then those ones originating away from
 the attachments. It follows that the creep events originating within
  the bending regions near attachment points contribute most of the creep noise; their contribution is
  greater by a factor of
  $\sim k^{-2}l^{-1}\lambda^{-1}=(l/\pi^2\lambda)(\omega_s/\omega)$ than that of the creep events away from the attachment points. This is studied in section 4.
 
  \subsection{The case of non-orthogonal laser beam and suspension fiber}
  Let us now consider the case where a laser beam  is inclined by a small angle $\beta$  with respect to
  the horizontal direction. This is an inevitable effect because of the spherical shape of the equipotential
  surface on which the test masses in the same arm are located; for $4$km arm $\beta\simeq 3\times 10^{-4}$ radians. It is this misalignment that was previously thought to be the major source of the creep noise \cite{cagnoli};
  we treat this mechanism within the formalism  developed in the previous section. The reciprocal
  force applied at the test mass has now a vertical component
  \begin{equation}
  F_{\rm vert}(t)=-\beta\Theta(t)
  \label{fvert}
  \end{equation}
  that causes the vertical test-mass motion
  \begin{equation}
  X_{\rm vert}=-{\beta l \Theta(t)\over N\pi E r^2}[1-\cos(\omega_{\rm vert}t)e^{-\gamma_{\rm vert}t}],
  \label{xvert}
  \end{equation}
  where the negative sign corresponds to the upward motion. Here we take into account only one vertical suspension mode with the angular frequency  
  \begin{equation}
  \omega_{\rm vert}=\sqrt{NE\pi r^2\over Ml}
  \label{omegavert}
  \end{equation}
  and the damping rate $\gamma_{\rm vert}$;
  the higher-order vertical modes are at much higher frequency and have a much weaker coupling to the sudden
  force. The readout variable from Eq.~(\ref{Xreadout2}) is given by
  \begin{equation}
  X_{\rm vertreadout}(t)={MgV\beta\Theta(t)\over E (N\pi r^2)^2}[\cos(\omega_{\rm vert}t)e^{-
  \gamma_{\rm vert}t}-1].
  \label{xvertreadout}
  \end{equation}
  In the Fourier domain,
  \begin{equation}
  X_{\rm vertreadout}(\omega)={\beta g V\over N\pi r^2 l}{1\over 2\pi i\omega
  [\omega_{\rm vert}^2-(\omega-i\gamma_{\rm vert})^2]}.
  \label{xverfourrier}
  \end{equation}
  It is instructive to compare the amplitude of the vertical mode to the amplitude of the pendulum mode
  excited by the creep even  near the attachment point, as inferred from Eq.~(\ref{xreadout}). Their ratio
  is approximately given by
  \begin{equation}
  {{\rm vertical}\over {\rm pendulum}}\simeq {\beta r^2\over x_c \lambda}\sim {\rm few}\times 10^{-5}.
  \label{verticalpendulum}
  \end{equation}
  It is the smallness of this ratio that makes the contribution to the creep noise from the creep-induced 
  fiber-lengthening be subdominant relative to the direct horizontal coupling, in most of the LIGO
  band.

 \begin{figure*}
  \includegraphics[scale=0.8]{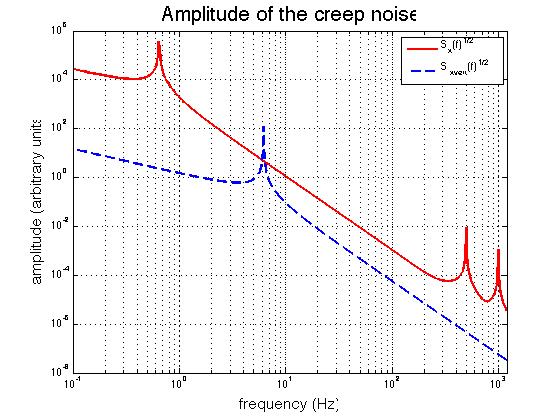}
\caption{\label{fig:epsart} Two amplitudes of the creep noise are plotted: that due to the direct horizontal
coupling (continuous), and that due to the vertical lengthening of the suspension fibers (dashed). While the units on the $y$-axis are arbitrary, the ratio of the two contributions depends on the elastodynamics only and is robust. One can see that the horizontal coupling makes dominant contribution everywhere except in a narrow band near the vertical resonance of the last stage of the suspension. The parameters for the plot are those for advanced LIGO suspension fused silica fibers: $E=72$GPa, $r=2\times 10^{-4}$m, $l=0.6$m, $M=40$kg, $N=4$. For this plot, the $Q$-factor of all the modes is taken to be $10^3$; this choice affects the height of the sharp peaks in the figure. Realistic $Q$ values will be several orders of magnitude higher.}
\end{figure*}

\section{IV. Creep noise in the stationary limit}
Consider now a situation where multiple creep events are triggered in sequence,. According to  the Central Limit Theorem, if the events occur sufficiently frequently, their superposition produces a random Gaussian
noise in the test-mass motion. The response of the test-mass to a single creep event
can be written as $X(\vec{\alpha}, t-t_0)$ where $t_0$ is the time when the creep event is 
triggered,  and $\vec{\alpha}$ is the set of parameters characterizing the event (location in the fiber,
effective volume, etc.). In what follows we assume that the creep events are statistically independent 
from each other and that the creep-event parameters  sample some well-defined probability-distribution function. This assumption is known not to hold  in some systems that exhibit so-called ``crackle noise" \cite{dahmen}, and will be relaxed in future work. If the probability density distribution $P(\vec{\alpha})$ is time-invariant, 
then the creep noise is stationary and has a spectral density given by 
\begin{equation}
S_X(f)=8\pi^2R\int d\vec{\alpha} |X(\vec{\alpha},\omega)|^2 P(\vec{\alpha}),
\label{specdensitycreep}
\end{equation}
where $R$ is the rate of the creep events. The $d\vec{\alpha}$ implies a multi-dimensional integral
over the parameter space of $\vec{\alpha}$.  Evaluating this expression for the model of the cylindrical
fiber, we get 
\begin{equation}
S_x(f)={2 R\langle V^2\rangle\over c_s^2} \left({g\over 2\pi N r}\right)^2\times  Q(\omega-i\gamma)G(k)
\label{sdensity}
\end{equation}
where 
\begin{equation}
Q(\omega)=\left|\pi{\omega\over\omega_s}\cos\left(\pi{\omega\over\omega_s}\right)\omega_p^2-\omega^2
\sin\left(\pi{\omega\over\omega_s}\right)\right|^{-2}
\label{Q}
\end{equation}
and
\begin{equation}
G(k)={1\over \lambda l}\left\{[1+\cos^2(kl)]+k\lambda[2\sin(2kl)+kl]\right\}.
\label{G}
\end{equation}
Here $\langle V^2\rangle$ is the ensemble average of the $(\hbox{volume})^2$ of the creep events in the system. 
Naturally, this quantity is meaningful only in our simple model for the creep events, however a term like this, representing the mean of the squared intensity of
the creep events, is expected in any generic model for the localised creep events. 
The first term in square brakets on the right-hand side of the above equations is due 
to creep events generated near the attachment points, while the second term is that due to creep events
generated in the fiber's bulk; in both of these the terms of order $k\lambda$ have been neglected. It is 
clear that the noise is strongly dominated by the creep events near the attachment point.

  The spectral sensity of noise due to the vertical lengthening of the fibers is given by
  \begin{equation}
  S_{\rm xvert}(f)={2R\langle V^2\rangle\beta^2\over \omega^2}\left({g\over N\pi r^2 l}\right)^2
  Q_{\rm vert}(\omega-i\gamma),
  \label{svert}
  \end{equation}
  where 
  \begin{equation}
  Q_{\rm vert}(\omega)=\left|\omega_{\rm vert}^2-\omega^2\right|^{-2}.
  \label{qvert}
  \end{equation}
  The plots for $\sqrt{S_x(f)}$ and $\sqrt{S_{\rm xvert}(f)}$ are shown in Fig.~2. While the scale on the
  vertical axis of these plots is arbitrary, since $R\langle V^2 \rangle$ is unknown, the spectral density 
  shape and the relative contribution of the two noises are fixed. We observe that the direct horizontal coupling induces greater creep noise than the vertical motion, at all frequencies except at a narrow
  band around $f=\omega_{\rm vert}/(2\pi)$.

\section{V. Discussion}
In this paper I have 
provided an elastodynamic calculation of the interferometer's response to a creep event, 
and found the functional form of the creep noise in the stationary limit. A simple model where
the fiber was modeled as a cylinder of constant radius was considered in detail, since this allowed me to
obtain analytical expressions for the test-mass response. 
Two interrelated qualitative features of this model are worth noting: (1) Creep events near the fiber's ends
receive a much stronger test-mass response in the LIGO band than those at the center of the fiber, and 
contribute the majority of the creep noise, and (2) the dominant coupling to the inteferometer's 
readout is via excitation of the violin and pendulum modes of the suspension, and not via the lenthening
of the fiber. I should caution though that these conclusions may not hold in a fiber with a more complex
dependence of the cross-sectional radius $r$ on the height $z$. In particular, the fibers in advanced LIGO
suspensions are made significantly thicker near the end points,  in order to minimize the suspension thermal noise. This thickening will reduce the local tension stress, thus reducing both the coupling of a
creep event to the test mass motion and the likelyhood of a creep event to occur.

In a simple model for the creep noise, 
I have assumed that the creep events are triggered homogeneously in the suspention fiber.
This may not be the case.  The creep events may be  triggered preferentially (1) at the locations 
where the fiber is welded to the test mass or the upper
 suspension plate, although this is not very likely since at the weld the fiber is much thicker than in its center ($r=1.5\times 10^{-3}$m), so the tension
 is small (Norna Robertson, private communications), or (2) near the locations where the ears that support the test mass are bonded to it.
 The bonding material is non-metallic and non-glassy and is a potential source of problems (Riccardo DeSalvo, private communications).
In future work I plan to explore the spatial distribution 
 of the expected creep events, as well as  relax the assumption of their statistical independence. I plan to
 also deal with the issues of non-Gaussianity   of the creep-event triggers;  it can presumably   can be mitigated by considering the output of several independent interferometers.
 
 Some comfort for the advanced interferometers can be derived from the fact that
 experiments \cite{bilenko2}, \cite{gretarsson}, and \cite{sorazu} have not observed any influence of the
 creep noise on the violin-mode motion. We note, however, that all of the measurements in question have searched for the creep noise at high frequencies corresponding to the resonant frequencies of violin modes, from several hundred to several thousand Hz. If the creep events are statistically
 independent from each other, then the expected creep noise is red, with 
 $\sqrt{S_X(f)}\propto f^{-3}$ except near resonances; see Eqs.~(\ref{sdensity}) and (\ref{Q}). This is 
 the same scaling as that for the suspension thermal noise in the case where  damping of the fiber's
 motion is structural (see, e.g., \cite{gonzales}). Therefore, one may argue that since no
  creep noise that exceeds the suspension thermal noise is observed at high frequencies, none is expected to exceed the suspension thermal noise at low frequencies as well.
 This argument, however,  relies on a very simple model for the creep noise that was developed
 in section 4, and in particular it relies on the creep events being statistically independent. This assumption does not hold in many systems that release their free energy via   spontaneous acoustic emission events (known as the ``crackle noise"), see \cite{dahmen} and references therein. Thus further experimental and theoretical work is warranted
 for the low-freqyency domain.

 I thank Vladimir Braginsky for impressing on me 
  the importance of non-stationary noise in many a conversation that we've had from the first time
  we met in September of 1994.  I thank  Misha Gorodetsky for his insightful comments given as part of the internal LSC review for this article. I thank Riccardo DeSalvo, Eric Gustafson, Jan Harms, Norna Robertson, and Peter Saulson for comments on the earlier version of this manuscript. I thank Rana Adhikari, Yanbei Chen,  Kip Thorne, and Sergei Vyatchanin for useful discussions. Finally, I thank  Sarah Levin for proofreading parts of the manuscript before submission. The calculations in this paper were completed during a visit to LIGO laboratory at Caltech. 
  The research was supported by the Australian Research Council Future Fellowship.

%\end{multicols}

  \appendix
\section{Appendix: Response of the suspension to a periodic force applied at the test mass}
Here we provide a quick derivation; similar derivations for more complicated situations
when the test-mass tilt is allowed is given in e.g.~Appendix A of \cite{braginsky2}. Suppose a periodic force 
\begin{equation}
F=F_0e^{i\omega t} 
\end{equation}
is acting on the test mass and induces its periodic motion
\begin{equation}
X_{\rm testmass}=X_0e^{i\omega t}.
\end{equation}
The fiber's motion is given by 
\begin{equation}
\eta(z,t)={\sin(kz)\over \sin(kl)}X_0 e^{i\omega t}.
\end{equation}
Here $k=\omega/c_s=(\pi/l)\omega/\omega_s$ is the tension-wave vector.
The horizontal component of back-reaction tension force acting on the test-mass is
\begin{equation}
F_{\rm fiber}=F_{f0}e^{i\omega t}=-Mgk\cot(kl)X_0e^{i\omega t}.
\label{A4}
\end{equation}
The second Newton's law gives
\begin{equation}
F_0+F_{f0}=-M\omega^2 X_0.
\end{equation}
Substituting the Eq.~(\ref{A4})  above we get
\begin{equation}
X_0={F_0\over M}{1\over gk\cot(kl)-\omega^2},
\end{equation}
which is equivalent to Eq.~(\ref{impedance}) in the text.

%%\section{Central Limit Theorem and spectral density of the creep noise}


\begin{thebibliography}{99}
\bibitem{footnote2}
Naturally, other time-dependencies may also be considered within this formalism.
\bibitem{footnote3}
More precisely, for frequencies such that the corresponding sound waves have
a much larger wavelength than the size of $V$, which is an excellent approximation in
our situation.
\bibitem{footnote4}
Strictly speaking,
this expression is valid only to zero'th order in $m/M$ at low frequencies, where $m$ and $M$ are  the
masses of the
string and the test mass, respectively. This is of order of $10^{-4}$ for LIGO.
\bibitem{footnote5}
This enforces causality of
the test-mass response. In case there is no damping, one needs to use the Landau rule when evaluating the inverse Fourier transform: the integration contour should pass under every pole.
\bibitem{footnote6}
In real suspension systems, this may not be so.
\end{thebibliography}
\end{document}